\documentclass[11pt]{article}
\usepackage[a4paper,margin=26mm]{geometry}
\usepackage[T1]{fontenc}
\usepackage{lmodern}
\usepackage{amsmath,amssymb,amsthm,mathtools}
\usepackage{microtype,cite}
\usepackage[hidelinks]{hyperref}
\newtheorem{theorem}{Theorem}[section]
\newtheorem{lemma}[theorem]{Lemma}
\newtheorem{corollary}[theorem]{Corollary}
\theoremstyle{definition}
\newtheorem{definition}[theorem]{Definition}
\theoremstyle{remark}
\newtheorem{remark}[theorem]{Remark}
\newcommand{\NoU}{\mathsf{N}}
\newcommand{\FA}{F_A}
\newcommand{\FB}{F_B}
\newcommand{\R}{R}
\newcommand{\out}{d^+}
\newcommand{\doi}[1]{\href{https://doi.org/#1}{\nolinkurl{doi:#1}}}
\hypersetup{
 pdftitle={Disjoint cycles of different lengths in 3-diregular digraphs with a two-cycle factor and girth at least five},
 pdfauthor={Duy Hieu Do},
 pdfsubject={Directed cycles in regular digraphs},
 pdfkeywords={directed cycles, cycle factors, regular digraphs, directed girth}
}
\title{Disjoint cycles of different lengths in 3-diregular digraphs\\
with a two-cycle factor and girth at least five}
\author{Duy Hieu Do\\[3pt]
\small Institute of Mathematics, Vietnam Academy of Science and Technology\\
\small Hanoi, Vietnam\\
\small\href{mailto:ddhieu@math.ac.vn}{\texttt{ddhieu@math.ac.vn}}}
\date{}

\begin{document}
\maketitle
\begin{abstract}
We prove that every finite simple digraph with indegree and outdegree
three at every vertex, directed girth at least five, and a spanning
cycle factor consisting of two cycles contains two vertex-disjoint
directed cycles of different lengths. The proof studies the strong
components remaining after a directed cycle is deleted. In a
counterexample, each cyclic component is a cycle of the deleted
cycle's length, and degree balance restricts the arcs between these
components. These restrictions yield a short cycle with acyclic
complement; a two-prefix argument on the remaining factor paths gives
a contradiction. Together with earlier results for girth four and
for factors with at least three components, the theorem gives the
same conclusion at girth at least four whenever a cycle factor has
at least two components. This combined girth bound is sharp.
\end{abstract}



\section{Introduction}

Thomassen~\cite{Thomassen1983} proved that minimum outdegree three
suffices for two vertex-disjoint directed cycles. For cycles of
different lengths, Henning and Yeo~\cite{HY2012} proved the conclusion
for 4-regular digraphs and conjectured it under minimum outdegree four;
Lichiardopol~\cite{Lichiardopol2014} established the latter statement.
In a recent preprint, Albrechtsen and Steiner~\cite{AS2026} proved
that, for every positive integer \(k\), sufficiently large minimum
outdegree forces \(k\) vertex-disjoint cycles of pairwise distinct lengths.

We consider 3-diregular digraphs, in which every vertex has indegree
and outdegree three. In this class the different-length conclusion
has exceptions. Tan~\cite{Tan2015} constructed an infinite family
with digons and no disjoint cycles of different lengths.
The girth-three case was classified in~\cite{Tan2017}, whereas every
3-diregular digraph of girth four has the desired pair~\cite{Tan2020}.
The latter work also proposed a broader classification of the
3-diregular exceptions. Related results have been obtained for
multipartite tournaments~\cite{HHT2022}.

Cycle factors provide a way to refine the degree-three problem.
Tan~\cite{Tan2015} proved the conclusion for bipartite 3-diregular
digraphs with a cycle factor having at least two components.
He, Cheng and Gao~\cite{HCG2022} treated factors with at least three
components, apart from a specified family of girth two.
Our main theorem treats a spanning factor with exactly two
components when the directed girth is at least five.

\begin{theorem}\label{thm:main}
Let \(D\) be a finite simple 3-diregular digraph of directed girth at least five.
If \(D\) has a spanning cycle factor with exactly two components, then
\(D\) contains two vertex-disjoint directed cycles of different lengths.
\end{theorem}

Combining Theorem~\ref{thm:main} with the girth-four and
three-component factor results gives the following consequence.

\begin{corollary}\label{cor:allfactors}
Let \(D\) be a finite simple 3-diregular digraph of directed girth at least four.
If \(D\) has a cycle factor with at least two components, then it has
two vertex-disjoint directed cycles of different lengths.
Consequently, in a 3-diregular digraph of girth at least four with no
such pair, every cycle factor is Hamiltonian.
\end{corollary}
\begin{proof}
If the girth is four, apply Tan's theorem~\cite{Tan2020}.
Suppose the girth is at least five. A factor with two components is
covered by Theorem~\ref{thm:main}. For a factor with at least three
components, apply He, Cheng and Gao~\cite{HCG2022}; their exceptional
digraphs contain digons and hence cannot occur here.
The final assertion is the contrapositive.
\end{proof}

\begin{remark}\label{rem:girth3}
The girth bound in Corollary~\ref{cor:allfactors} cannot be lowered
to three. The digraph \(D_8^3\) has vertices
\(v_0,\ldots,v_5,t,w\), arcs \(v_iv_{i+1}\) and \(v_iv_{i+2}\)
with indices modulo six, and arcs \(tv_i,v_iw\) for even \(i\)
and \(wv_i,v_it\) for odd \(i\).
It has girth three and no disjoint cycles of different lengths
by~\cite[Theorem~3]{Tan2017}. It nevertheless has the two-cycle factor
\[
 (v_0,v_1,v_3,v_4,v_0)\ \cup\ (v_2,w,v_5,t,v_2).
\]
\end{remark}

\subsection{Proof strategy}

Suppose that a counterexample exists. Its two factor cycles have
the same length \(m\), and every remaining arc joins the two cycles.
A two-run cycle consists of one nonempty interval of each factor
cycle and two joining arcs. A fundamental cycle is a two-run cycle
in which one interval is a singleton. Let \(h\) and \(\ell\) be their
respective minimum lengths, so \(h\le\ell\).

We first show that, after any directed cycle is deleted, each cyclic
strong component of its complement is a cycle of the same length.
This structural lemma uses strong connectivity and exact
3-diregularity, without a girth or factor assumption. For a two-run
cycle with cyclic complement, degree balance orders the cyclic
components into blocks separated by acyclic regions.
The original factor arcs give one path from each factor
cycle through every such region, with empty paths allowed.

Fundamental-cycle geometry yields \(m\ge2\ell-1\), whereas a
two-run cycle of length \(s\) with cyclic complement would give
\(m\le2s-2\). A shortest two-run cycle therefore has acyclic
complement. We process a maximal source prefix on one remaining
factor path and then a prefix on the other. The first prefix bounds
the predecessor indices available to the second. A predecessor
lemma then contradicts an arc forced by the first prefix.

The reduction and the two minima are introduced below.
Section~\ref{sec:components} proves the structural lemmas.
Section~\ref{sec:fundamental} gives the fundamental-cycle geometry
and numerical bound. Section~\ref{sec:reanchor} obtains the acyclic
complement, and Section~\ref{sec:prefix} completes the proof.
The proof of Theorem~\ref{thm:main} is self-contained; the external
factor and girth-four theorems are used only for
Corollary~\ref{cor:allfactors}.

\subsection{Notation and the two-cycle reduction}

All digraphs here are finite, have no loops, and have at most one arc
for each ordered pair of distinct vertices. Opposite arcs are allowed
in this definition, although the girth hypothesis excludes them.
A directed cycle is simple, and \(|C|\) denotes the number of vertices
of a cycle \(C\). A spanning cycle factor has exactly one
incoming and one outgoing arc at every vertex. A digraph is
\emph{3-diregular} if every indegree and every outdegree is three.
The directed girth is the length of a shortest directed cycle.
A strongly connected component is called \emph{cyclic} if it contains
a directed cycle. An acyclic digraph contains no directed cycle.

For a digraph \(G\), let \(V(G)\) and \(E(G)\) be its vertex and arc
sets, and let \(G[S]\) be the subdigraph induced by \(S\subseteq V(G)\).
We write \(G-S=G[V(G)\setminus S]\), \(G-x=G-\{x\}\), and
\(G-C=G-V(C)\) for a cycle \(C\).
The out- and in-neighbourhoods of \(x\) in \(G\) are
\(N_G^+(x)\) and \(N_G^-(x)\), and
\(d_G^\pm(x)=|N_G^\pm(x)|\).
For disjoint vertex sets \(P,Q\), \(e(P,Q)\) counts arcs from
\(P\) to \(Q\); a subdigraph argument is interpreted as its vertex set.
Strong components are compared by reachability. A strict predecessor
component of \(W\) is a different component that can reach \(W\),
and a strict successor is defined dually. A terminal component has
no outgoing arc to another component.

Throughout the proof suppose, for a contradiction, that \(D\) satisfies
the hypotheses and also
\[
 \NoU:\qquad
 \text{every two vertex-disjoint directed cycles have the same length}.
\]
The two components of the given factor must have the same length \(m\).
Label them
\[
 \FA=A_0A_1\cdots A_{m-1}A_0,\qquad
 \FB=B_0B_1\cdots B_{m-1}B_0.
\]
Put \(A=V(\FA)\) and \(B=V(\FB)\).
We also use \(F_A(x)\) and \(F_B(x)\) for the successors of \(x\)
on the corresponding factor cycles. Factor intervals include their
endpoints; a proper interval omits at least one cycle vertex.
The vertex set of the directed factor interval from \(x\) to \(y\)
is denoted by \([x,y]_{F_A}\) or \([x,y]_{F_B}\), with a singleton
when \(x=y\). Factor-cycle subscripts are taken modulo \(m\).
Inequalities between indices use their representatives in
\(\{0,\ldots,m-1\}\) after the stated normalization.
The notation \([i-j]_m\) means the least nonnegative residue modulo \(m\).
In particular \(m\ge5\).

There is no additional arc within either cycle. Such an arc would be a
chord, and its closure along that factor cycle would have length less than \(m\)
while avoiding the other factor cycle. Consequently the residual
digraph
\[
 \R=\bigl(V(D),E(D)\setminus(E(\FA)\cup E(\FB))\bigr)
\]
is bipartite between \(A\) and \(B\), with residual indegree and outdegree
two at every vertex.
For a subdigraph or vertex set \(Y\), write
\(A(Y)=A\cap V(Y)\) and \(B(Y)=B\cap V(Y)\), with \(V(Y)=Y\)
when \(Y\) is itself a vertex set.
All residual neighbourhoods are taken in this full residual digraph,
including neighbours later removed with a selected cycle.
Deleting factor arcs does not delete vertices. The digraph \(D\) is
strongly connected: each factor cycle is strongly connected and
residual arcs link the two cycles in both directions.

\begin{definition}
A \emph{fundamental cycle} has one vertex on one factor cycle, a nonempty
consecutive interval of the other factor cycle, and the two residual arcs
joining the ends. The singleton vertex is its \emph{marker}, and the
interval on the other factor cycle is its \emph{return interval}.
The cycle is called \emph{one-\(A\)} or \emph{one-\(B\)} according
to the marker's factor cycle, and \emph{proper} when its return interval
is proper. On a cycle meeting both \(A\) and \(B\), a \emph{run} is a
maximal consecutive segment following one original factor cycle.
A \emph{two-run cycle} has one nonempty consecutive interval of each
factor cycle, joined by two residual arcs. A run of one vertex contains
no factor arc. Let \(\ell\) be the minimum fundamental-cycle
length over both orientations, and let \(h\) be the minimum two-run
cycle length. These minima and all subsequent factor intervals refer
to the fixed original factor \(\FA\cup\FB\).
\end{definition}

\begin{lemma}\label{lem:fundbound}
The minima exist and satisfy
\[
 5\le h\le\ell\le m-1.
\]
\end{lemma}
\begin{proof}
Every residual path \(A_iB_jA_k\) closes along \(\FA\) to a fundamental
cycle. If all these cycles had length at least \(m\), their length
\([i-k]_m+2\) would force \([k-i]_m\in\{1,2\}\), where a return with
\(k=i\) has length two. Since \(B_j\) has two distinct residual
out-neighbours, whenever \(A_i\to B_j\) they would be exactly
\(\{A_{i+1},A_{i+2}\}\). Its second, distinct residual predecessor
\(A_t\) would give the same consecutive two-set with \(i\) replaced by
\(t\), impossible on a cycle of order \(m\ge3\).
Thus some fundamental cycle has length at most \(m-1\).
The other inequalities follow from the definitions and the girth.
\end{proof}

We shall use \(\NoU\) to prove both that a globally shortest two-run
cycle has acyclic complement and that it cannot have acyclic complement.

\section{Cycle complements and degree balance}\label{sec:components}

\begin{lemma}[Cyclic components after deleting a cycle]\label{lem:scc}
Let \(D\) be a finite simple strongly connected 3-diregular digraph
satisfying condition \(\NoU\).
Let \(C\) be any directed cycle of length \(s\), and put \(X=D-V(C)\).
Every cyclic strongly connected component of \(X\) is exactly a directed
\(s\)-cycle.
\end{lemma}
\begin{proof}
Every directed cycle in \(X\) has length \(s\), by \(\NoU\). Fix a cyclic
strong component \(W\). Every closed directed walk in \(W\) has length
divisible by \(s\), by decomposition into simple cycles. Fixing a root
therefore defines a cyclic partition
\[
 W=L_0\mathbin{\dot\cup}\cdots\mathbin{\dot\cup}L_{s-1},
 \qquad E(D[W])\subseteq\bigcup_{i=0}^{s-1} L_i\times L_{i+1}.
\]
Layer subscripts are taken modulo \(s\).
Indeed the lengths of two root-to-vertex paths are congruent modulo \(s\):
append the same path back to the root to both. Every vertex of \(W\)
lies on an \(s\)-cycle, so every layer occurs. Such a cycle meets each
layer once. In particular a vertex outside an \(s\)-cycle in \(W\) has
at most one predecessor on that cycle.

We first establish an entry barrier. Suppose
\(\out_{D[W]}(x)=1\) and \(K\) is a cycle in \(W-x\).
Every simple \(C\)-to-\(x\) path whose interior avoids \(C\) meets \(K\).
Suppose instead that such a path \(Q\), starting at \(c\in C\), avoids
\(K\). The vertex \(x\) has two distinct outgoing arcs outside \(W\).
If both heads lie on \(C\), close \(Q\) through each head and then along
\(C\) to \(c\). These are simple cycles with distinct lengths, because
the two heads have distinct directed distances to \(c\) on \(C\).
Each cycle is vertex-disjoint from the same cycle \(K\), so \(\NoU\)
would force both lengths to equal \(|K|=s\). This is a contradiction;
the two constructed cycles need not be disjoint from each other.

Otherwise an outside arc \(x\to f\) enters the union \(F\) of the strict
successor components of \(W\) in \(X\). This set is successor-closed in
\(X\). Choose a longest simple path in \(D[F]\) starting at \(f\), and
let \(z\) be its endpoint. Every out-neighbour of \(z\) in \(F\) lies
on this path. Each such arc closes a cycle in \(X\), hence a cycle of
length \(s\); only one position on the path can give that length.
Thus \(z\) has at most one out-neighbour in \(F\), and at least two
distinct out-neighbours on \(C\). The path \(Q\) avoids \(F\), since
a vertex of \(F\) cannot reach \(W\) in \(X\). Append \(x\to f\) and
the chosen path in \(F\) to \(Q\), and close through the two heads on
\(C\). The appended path is simple, since \(Q\) avoids \(F\), and
its interior avoids \(C\). Again the two resulting cycles have
distinct lengths and both avoid \(K\), giving the same contradiction.
This proves the barrier.

We next show that \(W-x\) is acyclic whenever
\(\out_{D[W]}(x)=1\). Suppose not. For each cycle \(K\subseteq W-x\), set
\[
 T_K=\{v\in W-V(K):v\text{ reaches }x
                     \text{ in }D[W-V(K)]\}.
\]
Length-zero paths are permitted, so \(x\in T_K\).
Every predecessor of \(T_K\) in the full digraph \(D\)
belongs to \(T_K\cup V(K)\).
For a predecessor in \(W-V(K)\) this follows by prepending its arc to
a path to \(x\). A predecessor \(u\notin W\) is impossible: choose a
simple path from \(C\) to \(u\), retaining the part after its last
\(C\)-vertex. If \(u\in C\), use the path of length zero. It avoids \(W\), since otherwise its suffix and the arc
back into \(T_K\subseteq W\) would put \(u\) in \(W\).
Appending the arc into \(T_K\) and a path to \(x\) contradicts the entry
barrier.

Choose \(K\) minimizing \(|T_K|\), and write \(T=T_K\).
Each vertex of \(T\) has at most one predecessor on \(K\), by the layer
partition. The full indegree of each vertex is three, and all other
predecessors lie in \(T\), so every vertex of \(D[T]\) has indegree
at least two. In particular \(|T|\ge3\), because the digraph is simple.
Deleting \(x\) removes at most one incoming arc at any remaining
vertex. Thus \(T-x\) is nonempty and has minimum indegree at least
one. Following incoming arcs in this finite digraph gives a cycle \(L\).
This cycle is disjoint from \(K\), avoids \(x\), and has length \(s\).

Every vertex of \(K\) is reachable from \(C\) by a path avoiding \(T\).
To see this, take a simple path from \(C\) to \(x\) with interior
outside \(C\). Just before its first entry into \(T\) it lies on \(K\).
Retain the part up to its first \(K\)-vertex and then follow \(K\)
to the required vertex.

No vertex of \(K\) can belong to \(T_L\). Otherwise the just constructed
path, followed by a path to \(x\) in \(W-V(L)\), would give a
\(C\)-to-\(x\) walk avoiding \(L\); deleting repetitions would contradict
the entry barrier for \(L\). Every intermediate vertex of a path witnessing membership in
\(T_L\) also belongs to \(T_L\). Hence that whole path avoids \(K\),
and
\[
 T_L\subseteq T_K.
\]
But \(L\subseteq T_K\) and \(L\cap T_L=\varnothing\), so this inclusion
is strict, contrary to the choice of \(K\).
Thus \(W-x\) is acyclic.

Finally, let \(x\) be the last vertex of a longest path in \(D[W]\).
Its internal outdegree is at least one by strong connectivity.
All its internal out-neighbours lie on the path, and two distinct
ones would close cycles of distinct lengths. Thus its internal
outdegree is one.
Every cycle of \(W\) contains \(x\). If another vertex occupied its
layer, a cycle through that vertex would contain two vertices in
the same layer, which is impossible. Thus this layer is a singleton.
Each vertex in the preceding layer then has internal outdegree one,
and hence belongs to every cycle by the preceding paragraph. An
\(s\)-cycle uses only one vertex from that layer, so it too is a
singleton. Continue backwards around the layers. All \(s\) layers
are singletons, and their permitted arcs give exactly one directed
\(s\)-cycle.
\end{proof}

\begin{lemma}[Capacity separator]\label{lem:separator}
Let \(C\) be a directed cycle of length \(s\), and suppose
\(D-C\) has a cyclic component \(W\). Let \(U\) be the union
of its strict predecessor components in \(D-C\), and set \(S=U\cup W\).
Then every component outside \(C\) is comparable with \(W\). All arcs
from \(C\) to \(D-C\) enter \(S\), and there are no arcs from \(U\)
to \(C\) or to \((D-C)-S\).
\end{lemma}
\begin{proof}
By Lemma~\ref{lem:scc}, \(W\) is an \(s\)-cycle.
The cycle \(C\) is induced: a chord would give a shorter cycle inside
\(C\), disjoint from \(W\). Both cycles thus have exactly \(2s\)
outgoing arcs to their complements.

There is no arc from \((D-C)-S\) into \(S\).
Balance of the exact indegrees and outdegrees over \(S\) gives
\[
 e(S,(D-C)-S)+e(S,C)=e(C,S)\le2s.
\]
All \(2s\) external outgoing arcs of \(W\) occur on the left:
none enters \(U\). Equality therefore holds throughout, and every
arc counted on the left has its tail in \(W\). This proves the stated
arc restrictions. For a vertex outside \(S\), take a path from \(C\)
to it with interior outside \(C\). Its first vertex enters \(S\)
and its first exit from \(S\) departs from \(W\). The remaining path
shows that this vertex is a strict successor of \(W\).
\end{proof}

\begin{corollary}[The block-and-gap model]\label{cor:chain}
Suppose a cycle \(C_0\), of length \(s\), consists of one factor
interval on each cycle and two residual arcs, and has cyclic complement.
Then \(C_0\) and the cyclic components outside it form a cyclic chain
\[
 C_0,T_0,C_1,T_1,\ldots,C_{L-1},T_{L-1},C_0,\qquad L\ge2.
\]
Call the cycles \(C_i\) \emph{blocks}, their cycle arcs \emph{block arcs},
and the intervening vertex sets \(T_i\) \emph{gaps}.
Every \(C_i\) is an induced \(s\)-cycle, and each \(T_i\) induces an
acyclic digraph. An arc other than a block arc stays in one gap, goes from a block
to its following gap or next block, or goes from that gap to the next
block. No arc goes backwards or skips a block.
Every block has one nonempty run on each original factor cycle, and each gap
has one factor path on each cycle, possibly empty. The block orders
on the original cycles agree.
\end{corollary}
\begin{proof}
Lemma~\ref{lem:separator} makes the cyclic components outside \(C_0\)
totally ordered by reachability and comparable with every other
component. Assign each remaining vertex to its region before, between
or after consecutive cyclic components. These regions are the gaps;
each induces an acyclic digraph, but components within one gap need
not be comparable.

The separator at an intervening cyclic component forbids an arc
jumping forward over it. Condensation order forbids backwards arcs
in \(D-C_0\). At the anchor boundary, the separator at the first
outside cyclic component confines all heads from \(C_0\) to the first
gap or that first component. The separator at the last component
allows arcs into \(C_0\) only from that component or the last gap.
An arc from the last gap to the first would already be backwards
in the condensation of \(D-C_0\).
These restrictions give all the asserted cyclic locality.

The anchor cuts both original factor cycles. If it covered an entire
one, its complement would be a proper subgraph of the other factor
cycle and would be acyclic. Every outside block is mixed: the only
cycle using vertices from just one original factor cycle would be
that whole factor cycle, which has been cut by \(C_0\).

For a block \(C_i\), let \(r_i\) be its number of \(A\)-runs, also
its number of \(B\)-runs. Since the block is induced, exactly \(r_i\)
arcs of \(\FA\) leave it and exactly \(r_i\) enter it.
Indeed an original factor arc joining two vertices of \(C_i\) must
be an arc of that induced cycle; each maximal \(A\)-run therefore
has precisely one entering and one leaving \(\FA\)-arc.

The restriction of \(\FA\) to \(A(T_i)\) is a union of maximal
directed paths. It has no cycle, since the gap is acyclic.
The original factor indegree and outdegree are one, so the first
vertex of each such path has a factor predecessor outside the gap,
and the last has a factor successor outside it. Locality places
these two vertices in \(C_i\) and \(C_{i+1}\), respectively, with
subscripts on blocks taken modulo \(L\).
Let \(p_i\) count these nonempty paths and let \(d_i\) count direct
\(\FA\)-arcs from \(C_i\) to \(C_{i+1}\). Every factor arc leaving
\(C_i\), or entering \(C_{i+1}\), belongs to exactly one of these
routes. Hence
\[
 r_i=p_i+d_i=r_{i+1}.
\]
Since \(r_0=1\), all \(r_i=1\).

There is consequently one route of \(\FA\) through each gap, with
a direct arc representing an empty route. Every \(A\)-vertex of the
gap lies on it: an additional path would need another entering
and leaving factor arc. The same argument applies to \(\FB\).
Following these routes proves that the orders of the blocks on
the two original factor cycles coincide.
\end{proof}

\begin{lemma}[The fundamental degree kernel]\label{lem:kernel}
Let \(K\) be a fundamental cycle with marker \(b\in B\) and a
nonempty proper \(A\)-interval \(I\). Let \(b^-\) be the factor
predecessor of \(b\), and let \(a^-\) precede the first vertex of \(I\).
Put \(r=|N^+_{\R}(b^-)\setminus I|\). Then \(r\ne2\).
If \(r=1\), its unique surviving target is \(a^-\), the arc
\(a^-\to b\) is present, and \(D-K\) has a terminal component consisting
of \(a^-\) and \(|I|\) consecutive \(B\)-vertices ending at \(b^-\).
The reversed assertions hold for residual predecessors at the
factor successor of a marker. No minimum-length or induced-cycle
hypothesis is imposed on \(K\); its return interval is required
to be nonempty and proper.
\end{lemma}
\begin{proof}
In \(G=D-K\), every remaining \(A\)-vertex has outdegree at least two
except possibly \(a^-\). This vertex has outdegree one only if
\(a^-\to b\) was removed. Every \(B\)-vertex other than \(b^-\) retains
its factor arc, and has outdegree one only if both residual targets
were removed. The vertex \(b^-\) has outdegree \(r\).

If \(r\ge1\), every vertex of \(G\) has outdegree at least one.
Every terminal strong component is therefore cyclic, and hence a
directed cycle by Lemma~\ref{lem:scc}. Each of its vertices has
full outdegree one in \(G\), not merely outdegree one inside that
component. The degree-one \(B\)-vertices other than \(b^-\)
have only their factor arcs; these form a subgraph of the \(B\)-path
obtained by deleting \(b\). Its only possible return uses \(b^-\)
and \(a^-\). If \(r=2\), \(b^-\) is not degree one, so no return exists,
a contradiction. If \(r=1\), the return forces the surviving arc
\(b^-\to a^-\), the removed arc \(a^-\to b\), and a consecutive
\(B\)-path ending at \(b^-\). Its length is \(|K|\), so the path has
\(|I|\) vertices. There is only one possible terminal component since
all would contain \(a^-\) and \(b^-\).
Reverse all arcs for the dual statement.
\end{proof}

We will repeatedly use the following explicit form of the dual kernel.
Suppose a proper one-\(A\) fundamental cycle uses
\(B_i\to u\to B_k\) and the factor interval from \(B_k\) to \(B_i\).
If \(u\to B_{i+1}\) is absent, then
\begin{equation}\label{eq:sourcekernel}
 N^-_{\R}(F_A(u))\subseteq[B_k,B_i]_{F_B}.
\end{equation}
Indeed two predecessors outside that interval are forbidden by the
dual kernel; one outside predecessor would force the missing arc.
The \(A,B\)-interchanged implication also holds.

\section{Fundamental cycles and a length bound}\label{sec:fundamental}

\begin{lemma}[Shared-marker exclusion]\label{lem:bowtie}
Two proper fundamental cycles with the same \(B\)-marker cannot have
disjoint \(A\)-intervals.
\end{lemma}
\begin{proof}
Call the intervals \(I,J\), and let \(b^-\) precede the marker \(b\).
The kernel forbids both targets of \(b^-\) from lying outside either
interval. If both lie in \(I\), both lie outside \(J\), and conversely.
Thus there is exactly one target in each, and both kernel applications
give the terminal cycles supplied by Lemma~\ref{lem:kernel}.

Each such cycle contains the last two \(B\)-vertices before \(b\):
its \(B\)-interval has \(|I|\ge4\), respectively \(|J|\ge4\), vertices
by the girth hypothesis. At the earlier of these two \(B\)-vertices,
the factor arc to the later one is the unique outgoing arc in
both complements. Its two residual targets must therefore be deleted
in both complements, so must belong to \(I\cap J\), impossible.
\end{proof}

\begin{lemma}[One-marker block geometry]\label{lem:marker}
Let \(C_0\) be a proper one-\(B\) fundamental cycle whose complement
is cyclic. Its block-and-gap model has \(L=2\).
The extra residual in-tail and out-head at its marker lie in the
other block's \(A\)-run, in that strict order.
They are respectively the first and last \(A\)-vertices of that block,
and both runs of that block have at least two vertices.
\end{lemma}
\begin{proof}
The outside cycle makes \(C_0\) induced. Normalize its \(A\)-interval as
\(A_0,\ldots,A_{q-1}\), and write the two extra marker arcs as
\(A_t\to b\to A_s\), with \(q\le s,t<m\).
A digon excludes equality. If \(s<t\), the interval from \(A_s\)
to \(A_t\) closes a second fundamental cycle through \(b\), disjoint
from the original interval, contradicting Lemma~\ref{lem:bowtie}.
Thus \(t<s\) in the linear outside factor order.

If \(L\ge3\), locality places the out-head in the following gap or
next block and the in-tail in the preceding gap or preceding block.
The next and preceding blocks are distinct and both have nonempty
\(A\)-runs. This gives \(s<t\), impossible.
Hence \(L=2\). In this case the same order and locality put both
endpoints in the other block \(C_1\), with \(A_t\) before \(A_s\).
In particular its \(A\)-run has at least two vertices.

Suppose \(A_s\) is not the last vertex of that run. Its factor successor
\(x\) is still in \(C_1\) and is not first. As \(C_1\) is induced,
it has no residual predecessor inside \(C_1\): its incoming cycle
arc is already its factor arc. Locality permits external predecessors
only in \(B(C_0)\cup B(T_0)\). But \(B(C_0)=\{b\}\), and the two
residual heads of \(b\) are already the first vertex of \(C_0\) and
\(A_s\). Thus both residual predecessors of \(x\) lie in \(B(T_0)\).

The vertex \(A_s\) is not last, so its outgoing cycle arc is also
its factor arc. Inducedness excludes internal residual heads.
Locality puts its residual heads in \(B(T_1)\cup\{b\}\), and a head
\(b\) would form a digon with \(b\to A_s\).
Choose an external residual head \(w\in B(T_1)\).
The one-\(A\) fundamental cycle \(bA_swF_B\cdots b\) avoids \(B(T_0)\).
Its \(B\)-interval is proper because it omits the nonempty
\(B\)-run of \(C_1\).
At its marker successor \(x\), two residual predecessors remain outside
its \(B\)-interval, contradicting the dual kernel.
Thus \(A_s\) is last. Reversal makes \(A_t\) first.
If the \(B\)-run of \(C_1\) were a singleton \(b_1\), the four arcs
\(bA_sb_1A_tb\) would form a quadrangle, contrary to the girth.
\end{proof}

This argument used no minimum length for \(C_0\).
It also applies when \(|C_0|=m\): the cycle deletes vertices from
both original cycles, so the mixed-block and run-conservation arguments
in Corollary~\ref{cor:chain} still apply.

\begin{lemma}\label{lem:numeric}
The minimum fundamental length satisfies
\[
 m\ge2\ell-1.
\]
\end{lemma}
\begin{proof}
Choose a shortest fundamental cycle over both orientations, interchanging
\(A\) and \(B\) so that
\(C_0=A_0\cdots A_{\ell-2}bA_0\).
If its complement is cyclic, Lemma~\ref{lem:marker} leaves one other
block \(C_1\). Write its run sizes as \(a,b_1\), where \(b_1\ge2\).
Set \(v_i=|B(T_i)|\) for \(i=0,1\), where \(T_0\) follows
\(C_0\) and \(T_1\) precedes \(C_0\).

Let \(f,g\) be the first and last \(A\)-vertices of \(C_1\).
They are distinct. The vertex \(f\) has one internal residual
predecessor and another predecessor outside \(C_1\).
The latter is not the marker of \(C_0\), whose two targets are
the first \(A\)-vertex of \(C_0\) and \(g\). Locality therefore puts
it in \(B(T_0)\). With the arc from \(f\) to that marker, it gives a
one-\(A\) fundamental cycle of length at most \(v_0+2\).
Likewise \(g\) has one internal residual head and an external one.
An external head equal to the marker would create a digon, so it
lies in \(B(T_1)\), giving a fundamental cycle of length at most
\(v_1+2\).
Minimality on both cycles yields \(v_i\ge\ell-2\). Hence
\[
 m=1+b_1+v_0+v_1\ge1+2+2(\ell-2)=2\ell-1.
\]

Otherwise \(X=D-C_0\) is acyclic. Write \(b=B_0\) and the extra marker
arcs as \(A_t\to b\to A_s\). The shortest fundamental
cycle \(C_0\) is induced, since any additional residual arc through
its marker would close a shorter fundamental cycle. Thus both extra
endpoints are outside its \(A\)-interval. The shared-marker exclusion
in Lemma~\ref{lem:bowtie} again gives \(t<s\) in the linear outside
\(A\)-factor order.
Fix a topological ordering of \(X\), and write \(x\prec y\)
when \(x\) precedes \(y\) in this order.
Put \(u=m-\ell+1\).
Every residual predecessor \(B_i\) of \(A_t\) has
\(i+2\ge\ell\), by its one-\(A\) fundamental return to \(b\).
Here \(i\ne0\), since \(B_0\to A_t\to B_0\) would be a digon.
The two distinct such indices have maximum at least \(\ell-1\).
They precede \(A_t\) in the acyclic complement, so
\(B_{\ell-1}\prec A_t\) in its topological order.
Dually, a residual head \(B_j\) of \(A_s\) satisfies
\(m-j+2\ge\ell\), or \(j\le u+1\).
Here \(j\ge1\), since \(A_s\to B_0\) would be a digon with
the extra arc \(B_0\to A_s\).
The smaller of its two indices is at most \(u\), and
\(A_s\prec B_u\). The outside \(A\)-path puts \(A_t\prec A_s\).
Consequently
\[
 B_{\ell-1}\prec A_t\prec A_s\prec B_u .
\]
The \(B\)-vertices retain factor order in \(X\), so \(u\ge\ell\).
This is \(m\ge2\ell-1\).
\end{proof}

\section{A short cycle with acyclic complement}\label{sec:reanchor}

\begin{lemma}\label{lem:opposite}
A one-\(B\) fundamental cycle and a one-\(A\) fundamental cycle
cannot be vertex-disjoint.
\end{lemma}
\begin{proof}
If they were, \(\NoU\) would give equal lengths. Both intervals are
proper: the \(A\)-interval of the one-\(B\) cycle omits the other
cycle's \(A\)-marker, and the \(B\)-interval of the one-\(A\) cycle
omits the first cycle's \(B\)-marker.
Anchor the block model at the one-\(B\) cycle.
The complementary cyclic component containing
the one-\(A\) cycle is exactly that cycle, by Lemma~\ref{lem:scc}.
Lemma~\ref{lem:marker} gives just one other block and puts the two
distinct extra marker endpoints in its \(A\)-run. But that run has
one vertex. This is impossible.
\end{proof}

\begin{lemma}\label{lem:tworuncyclic}
If a two-run cycle \(C\) of length \(s\) has cyclic complement, then
\[
 m\le2s-2.
\]
\end{lemma}
\begin{proof}
The cycle \(C=C_0\) is induced, since a chord would yield a shorter
cycle disjoint from an outside \(s\)-cycle.
Use Corollary~\ref{cor:chain}.
If a run covered a whole original factor cycle, its complement would be a
proper subgraph of the other factor cycle and hence acyclic.
Thus both runs are proper.

An \(A\)-escape is a fundamental cycle whose unique \(B\)-marker is
in \(B(C)\) and whose \(A\)-interval is wholly outside \(A(C)\).
A \(B\)-escape is defined by interchanging \(A\) and \(B\).
Such opposite escapes are vertex-disjoint, so cannot coexist by
Lemma~\ref{lem:opposite}.

If there were at least three blocks, any \(B\)-vertex of \(C\)
would have an external residual predecessor in the preceding
block or gap and an external successor in the following gap or block.
These incidences exist because residual degrees are two and an
induced two-run cycle uses at most one in-arc or out-arc there.
In the linear outside \(A\)-order the successor precedes the predecessor.
Closing through that marker gives an \(A\)-escape.
Interchanging \(A\) and \(B\) gives a \(B\)-escape, a contradiction.
Thus there are exactly two blocks, \(C,C_1\), with gaps \(T_0,T_1\).

Suppose that \(A\)-escapes are absent. At any \(B\)-vertex of \(C\),
each external predecessor \(p\) and external successor \(q\) lie
respectively in \(A(C_1)\cup A(T_1)\) and
\(A(T_0)\cup A(C_1)\). If \(q\) precedes \(p\), they give an
\(A\)-escape; equality is a digon. Hence \(p\) precedes \(q\)
for every such pair. In the linear order outside \(A(C)\), the regions
occur as \(A(T_0),A(C_1),A(T_1)\). Thus all these external endpoints
lie in \(A(C_1)\). In particular no \(B\)-vertex of \(C\) targets \(A(T_0)\).

If \(T_0\) has an \(A\)-vertex, let \(x\) be its first one.
Choose a topological order of \(D[T_0]\), and take the \(B\)-vertices
preceding \(x\). They form an initial segment of the factor path
on \(B(T_0)\), because that path retains all its factor arcs.
Both residual predecessors of \(x\) lie in this segment: locality
permits them only in \(B(C)\cup B(T_0)\), and the absence of
\(A\)-escapes has already excluded heads from \(B(C)\) into \(A(T_0)\).
In particular this segment has at least two vertices.

Each vertex in the segment has both residual predecessors in
\(A(C)\). A predecessor in \(A(T_0)\) would precede \(x\), whereas
every vertex of the \(A\)-factor path in the gap is either \(x\) or
follows it. No other predecessor is allowed by locality.
Both residual heads of a segment vertex lie outside \(A(C)\), since
an arc from \(T_0\) back to \(C\) is forbidden.
Normalize \(A(C)=\{A_0,\ldots,A_{q-1}\}\), so \(x=A_q\),
and let \(O_j\subseteq\{q,\ldots,m-1\}\) be the two outside
target indices of prefix vertex \(B_j\).

For consecutive prefix vertices,
\begin{equation}\label{eq:headorder}
 \min O_{j-1}\ge\max O_j.
\end{equation}
Otherwise take a head \(A_c\) of \(B_{j-1}\) and a head \(A_k\) of
\(B_j\) with \(q\le c<k\), and the smaller predecessor \(A_i\) of \(B_j\).
If \(q=1\), its two distinct predecessors in \(A(C)\) are already
impossible. Otherwise \(i\le q-2\).
The fundamental return interval through \(A_iB_jA_k\) is
\([A_k,A_i]_{F_A}=\{A_k,\ldots,A_{m-1},A_0,\ldots,A_i\}\),
a proper interval.
Indeed it omits \(A_c\), since \(i\le q-2<c<k\).
Thus the head \(A_c\) of \(B_{j-1}\) survives when this cycle is
deleted. If its other head also survives, the kernel gives a
contradiction immediately. Otherwise the kernel forces the unique
surviving head to be \(A_{k-1}\), and also forces
\(A_{k-1}\to B_j\). But \(k-1\ge q\), whereas both residual
predecessors of \(B_j\) belong to \(A(C)\). This is again impossible.
This proves \eqref{eq:headorder}.
Every target set has two distinct indices at least \(q\), so only
the last prefix vertex can target \(x\): if a nonlast set contained
\(q\), the next set would have maximum at most \(q\), which cannot
hold for two distinct indices both at least \(q\).
This contradicts the two distinct residual predecessors of \(x\).

Thus \(T_0\) has no \(A\)-vertices. Reversal, which preserves absence
of \(A\)-escapes as a statement about cycle vertex sets, gives the
same for \(T_1\). If \(B\)-escapes rather than \(A\)-escapes are absent,
the exchanged argument empties the two \(B\)-gaps.
At least one escape type is absent. On that factor cycle the two block runs
cover all \(m\) vertices. Each mixed \(s\)-cycle has at most \(s-1\)
vertices there, so \(m\le2s-2\).
\end{proof}

\begin{corollary}\label{cor:dag}
Every two-run cycle of length at most \(\ell\), and in particular
every globally shortest two-run cycle, has acyclic complement.
\end{corollary}
\begin{proof}
A cyclic complement at length \(s\le\ell\) would give
\(m\le2s-2\le2\ell-2\), contrary to Lemma~\ref{lem:numeric}.
\end{proof}

\section{Two prefixes and the final contradiction}\label{sec:prefix}

Fix a globally shortest two-run cycle and normalize it as
\[
 C=A_0\cdots A_{a-1} B_0\cdots B_{b-1}A_0,\qquad
 a,b\ge1,\quad a+b=h.
\]
It is induced. Any additional \(A_i\to B_j\) closes with the selected
return \(B_{b-1}\to A_0\) to a shorter two-run cycle, unless it is the
existing arc \(A_{a-1}\to B_0\). The other direction is identical.
Thus exactly \(2a-1\) residual arcs leave its \(A\)-run for outside
\(B\), and exactly \(2b-1\) leave its \(B\)-run for outside \(A\).
By Corollary~\ref{cor:dag}, \(X=D-C\) is acyclic.
In this section a head is called \emph{old} when it belongs to the
corresponding run of \(C\). All neighbourhoods still refer to the
full residual digraph \(\R\), rather than to \(X\).

\begin{lemma}[A fixed range of predecessor indices]\label{lem:frozen}
Let \(1\le t\le m-a\), and consider the initial segment
\[
 x_r=A_{a+r-1}\qquad(1\le r\le t)
\]
of the outside \(A\)-factor path. All indices in this lemma are
linear representatives in \(\{0,\ldots,m-1\}\).
For each vertex \(x_r\), define its full residual predecessor-index set by
\[
 I_r=\{j:B_j\to x_r\in E(\R)\}.
\]
It has two elements and records every residual predecessor, including
those deleted with \(C\).
Suppose an integer \(\mu\) satisfies
\[
 b-1\le \mu\le h-2,
\]
and for every vertex in this prefix both actual residual predecessor
indices belong to \(\{0,\ldots,\mu\}\), while every residual head
outside \(C\)'s \(B\)-run has index greater than \(\mu\).
Then every head is outside the \(B\)-run, and
\begin{equation}\label{eq:descent}
 \max I_{r+1}\le\min I_r\qquad(1\le r<t).
\end{equation}
The indices in \(I_r\) may correspond to vertices deleted with \(C\)
or to previously processed outside vertices.
\end{lemma}
\begin{proof}
The existence of two distinct inputs in the nonempty prefix implies
\(\mu\ge1\). If \(\mu=0\), the hypotheses already contradict each other.
Suppose \(x_r\) is the first vertex to have an old head \(B_c\), with
\(0\le c<b\).
For an earlier vertex \(x_j\), put \(i=\min I_j\le \mu-1\) and choose
an outside head \(B_k\), so \(k>\mu\).
Since \(i+1\le\mu<k\), the fundamental return interval
for \(B_ix_jB_k\) is proper and is
\[
 [B_k,B_i]_{F_B}=\{B_k,\ldots,B_{m-1},B_0,\ldots,B_i\}.
\]
The arc \(x_j\to B_{i+1}\) is absent, since both actual heads have
index greater than \(\mu\). By \eqref{eq:sourcekernel}, every predecessor
\(B_v\) with \(v\in I_{j+1}\) lies in this interval. Since
\(v\le\mu<k\), it follows that \(v\le i\).
Thus \(\max I_{j+1}\le\min I_j\). This holds up to \(x_r\).
More precisely, it is applied for \(1\le j<r\), where
\(F_A(x_j)=x_{j+1}\) is still an outside vertex. It therefore bounds
the inputs of \(x_r\) without making any assumption about that
vertex's heads. No successor beyond the prescribed prefix is used.

If \(c=0\), the smaller input \(i\le \mu-1\) in \(I_r\) gives the
fundamental cycle \(B_0\cdots B_ix_rB_0\), of length
\(i+2\le \mu+1\le h-1\), impossible.
For \(c\ge1\), suppose first that \(c>r\). The arc
\(x_r\to B_c\), the selected return \(B_{b-1}\to A_0\), and
the two factor intervals form the cycle
\[
 A_0\cdots A_{a+r-1} B_c\cdots B_{b-1} A_0
\]
of length \(h+r-c<h\). This is a simple two-run cycle.
Its \(A\)-interval is proper because
\(r\le c-1\le b-2\) gives \(a+r\le h-2<m\); its
\(B\)-interval has at most \(b<m\) vertices.
Minimality of \(h\) therefore gives \(c\le r\).

Set \(j_0=r-c+1\). The \(c\) two-element input sets
\(I_{j_0},\ldots,I_r\) obey the descent inequality just proved.
If \(I_j=\{u_j,v_j\}\), where \(u_j<v_j\), it says
\(v_{j+1}\le u_j\le v_j-1\). Since \(v_r\ge1\), we have
\(v_{j_0}\ge v_r+c-1\ge c\). Choose \(i=v_{j_0}\in I_{j_0}\).
Now
\[
 B_c\cdots B_i x_{j_0}x_{j_0+1}\cdots x_r B_c
\]
has the two joining arcs given by the actual input
\(B_i\to x_{j_0}\) and the assumed head \(x_r\to B_c\).
The \(A\)-vertices are distinct consecutive vertices of the outside
factor path. The \(B\)-vertices are also distinct, since
\(c\le i\le\mu<m\). Thus it is a simple two-run cycle, with
\(i-c+1\) vertices on \(B\) and exactly \(c\) on \(A\).
Both intervals are proper: \(c<b<m\), and
\(i-c+1\le\mu<h<m\).
Its length is \(i+1\le\mu+1\le h-1\), a contradiction.
This cycle is constructed in \(D\); it may
use vertices deleted with \(C\), or vertices processed earlier.
Neither possibility affects the definition of \(h\).
Thus there is no old head.
The same source-kernel argument for every consecutive pair gives
\eqref{eq:descent}.
\end{proof}

\begin{remark}
The cutoff \(\mu\) need not lie inside the removed \(B\)-run.
When \(\mu\ge b\), the sets \(I_r\) may include indices of vertices
already processed on the outside \(B\)-path. Only the common index bound
is used in the lemma.
\end{remark}

\begin{proof}[Proof of Theorem~\ref{thm:main}]
We apply Lemma~\ref{lem:frozen} twice. The second application includes
the input vertices made available by the first prefix.
Every source of \(X\) is one of its first factor-path vertices,
\(A_a\) or \(B_b\), since every other vertex retains its factor
predecessor. Interchange \(A\) and \(B\) if necessary so that \(B_b\) is a source.
Process \(B_b\) and then consecutive \(B\)-vertices for as long as
possible, without processing an outside \(A\)-vertex. Here processing
means successively removing sources of the remaining acyclic digraph.
Such a sequence is an initial segment of a topological order.
This gives a
maximal nonempty initial prefix
\[
 B_b,\ldots,B_M,\qquad p=M-b+1\ge1 .
\]
Each has both residual predecessors in \(C\)'s \(A\)-run.
The available \(2a-1\) residual arcs give
\begin{equation}\label{eq:capacity}
 2p\le2a-1,\qquad p\le a-1,\qquad M\le h-2 .
\end{equation}
If \(a=1\), this is already impossible. Otherwise \(a\ge2\).
The outside \(B\)-path has \(m-b\ge a+1\) vertices by
\(m\ge h+1\), so this prefix cannot exhaust it.
Its successor \(z=B_{M+1}\) is blocked at the end of the prefix.
Since its factor predecessor has been processed, it has a residual
predecessor outside \(C\)'s \(A\)-run.

Apply the \(A,B\)-interchanged form of Lemma~\ref{lem:frozen} to the first
prefix with opposite cutoff \(\mu=a-1\). This is at most \(h-2\)
because \(b\ge1\). Its hypotheses hold: its input
indices lie in \(\{0,\ldots,a-1\}\), and any outside \(A\)-head has
index at least \(a\). Thus each first-prefix vertex has two outside
\(A\)-heads. Let \(A_k\) be the larger head of \(B_M\); then
\begin{equation}\label{eq:k}
 k\ge a+1.
\end{equation}
Choose its smaller residual input \(A_i\), where \(i\le a-2\).
The proper fundamental cycle through \(A_iB_MA_k\) has return interval
\[
 [A_k,A_i]_{F_A}=\{A_k,\ldots,A_{m-1},A_0,\ldots,A_i\}.
\]
The arc \(B_M\to A_{i+1}\) is absent since its heads are outside \(C\).
Applying the \(A,B\)-interchanged form of \eqref{eq:sourcekernel}
at \(z=F_B(B_M)\) puts all its residual predecessors in that return
interval. Its outside predecessor, say \(A_T\), therefore has
\(T\ge k\).

In any topological extension of the first prefix, all of
\(A_a,\ldots,A_T\) precede \(z\), by their factor path to \(A_T\to z\).
No vertex \(B_j\) with \(j\ge M+1\) can be a predecessor of one
of these vertices: the factor path from \(z\) to \(B_j\) would
put \(z\) both before and after that vertex. Equivalently, an arc
\(B_j\to A_u\), with \(a\le u\le T\), would close the outside
directed cycle
\[
 B_{M+1}\cdots B_j A_u\cdots A_T B_{M+1},
\]
where the first factor segment is a singleton when \(j=M+1\).
Thus their actual residual input indices are all at most \(M\).

Every residual head from this \(A\)-prefix to outside \(C\)'s
\(B\)-run has index greater than \(M\). To verify this independently
of any chosen topological extension, each processed vertex
\(B_j\), \(b\le j\le M\), already has its two residual predecessors
in the deleted \(A\)-run. Residual indegree is exactly two, so it
cannot receive an additional arc from an outside \(A\)-vertex.
This proves the required output cutoff.

This second \(A\)-prefix therefore satisfies
Lemma~\ref{lem:frozen}, with cutoff \(\mu=M\), using the bound from
\eqref{eq:capacity}. Its first vertex has smaller input at most \(M-1\).
By \eqref{eq:descent}, every vertex after the first has both inputs
at most \(M-1\).
But \(A_k\) belongs to this prefix, is after its first vertex by
\eqref{eq:k}, and has the actual predecessor \(B_M\).
This is a contradiction.

We have contradicted the acyclicity guaranteed by
Corollary~\ref{cor:dag}. Hence \(\NoU\) is false, proving
Theorem~\ref{thm:main}.
\end{proof}

\bigskip
\noindent\textbf{Funding.}
No external funding was received for this work.

\smallskip
\noindent\textbf{Competing interests.}
The author declares no competing interests.

\smallskip
\noindent\textbf{Data availability.}
This article has no associated datasets.

\smallskip
\noindent\textbf{Author contributions.}
The author is responsible for the mathematical arguments and the
preparation of the manuscript.


\begin{thebibliography}{99}
\small
\setlength{\itemsep}{0pt}

\bibitem{Thomassen1983}
C.~Thomassen,
Disjoint cycles in digraphs,
\emph{Combinatorica} \textbf{3}(3--4) (1983), 393--396.
\doi{10.1007/BF02579195}.

\bibitem{HY2012}
M.~A.~Henning and A.~Yeo,
Vertex disjoint cycles of different length in digraphs,
\emph{SIAM Journal on Discrete Mathematics}
\textbf{26}(2) (2012), 687--694.
\doi{10.1137/100802463}.

\bibitem{Lichiardopol2014}
N.~Lichiardopol,
Proof of a conjecture of Henning and Yeo on vertex-disjoint directed cycles,
\emph{SIAM Journal on Discrete Mathematics}
\textbf{28}(3) (2014), 1618--1627.
\doi{10.1137/130922653}.

\bibitem{AS2026}
S.~Albrechtsen and R.~Steiner,
Proof of Lichiardopol's conjecture on disjoint directed cycles of distinct lengths,
\emph{arXiv preprint} arXiv:2608.20012 (2026).
\doi{10.48550/arXiv.2608.20012}.

\bibitem{Tan2015}
N.~D.~Tan,
On vertex disjoint cycles of different lengths in 3-regular digraphs,
\emph{Discrete Mathematics} \textbf{338}(12) (2015), 2485--2491.
\doi{10.1016/j.disc.2015.06.016}.

\bibitem{Tan2017}
N.~D.~Tan,
On 3-regular digraphs without vertex disjoint cycles of different lengths,
\emph{Discrete Mathematics} \textbf{340}(8) (2017), 1933--1943.
\doi{10.1016/j.disc.2017.03.024}.

\bibitem{Tan2020}
N.~D.~Tan,
On 3-regular digraphs of girth 4,
\emph{Discrete Mathematics} \textbf{343}(1) (2020), Article 111632.
\doi{10.1016/j.disc.2019.111632}.

\bibitem{HHT2022}
L.~X.~Hung, D.~D.~Hieu and N.~D.~Tan,
Vertex-disjoint cycles of different lengths in multipartite tournaments,
\emph{Discrete Mathematics} \textbf{345}(6) (2022), Article 112819.
\doi{10.1016/j.disc.2022.112819}.

\bibitem{HCG2022}
Z.~He, P.~Cheng and Y.~Gao,
Disjoint cycles of different lengths in 3-regular digraphs,
\emph{Graphs and Combinatorics} \textbf{38}(3) (2022), Article 74.
\doi{10.1007/s00373-022-02465-3}.

\end{thebibliography}
\end{document}